\documentclass{emulateapj}
\usepackage{epsf}
\usepackage{xspace}
\usepackage{amsmath}
\usepackage{framed} 
\def\euv{\epsilon_{\rm UV}}

\def\dim#1{\mbox{\,#1}}

\def\hide#1{}

\begin{document}

\title{Cosmic Reionization On Computers. Properties of the Post-reionization IGM}

\author{Nickolay Y.\ Gnedin\altaffilmark{1,2,3}, George D.\ Becker\altaffilmark{4},  Xiaohui Fan\altaffilmark{5}}
\altaffiltext{1}{Particle Astrophysics Center, Fermi National Accelerator Laboratory, Batavia, IL 60510, USA; gnedin@fnal.gov}
\altaffiltext{2}{Kavli Institute for Cosmological Physics, The University of Chicago, Chicago, IL 60637 USA}
\altaffiltext{3}{Department of Astronomy \& Astrophysics, The University of Chicago, Chicago, IL 60637 USA} 
\altaffiltext{4}{Space Telescope Science Institute, 3700 San Martin Dr, Baltimore, MD 21218, USA}
\altaffiltext{5}{Steward Observatory, University of Arizona, Tucson, AZ 85721}

\begin{abstract}
We present a comparison between several observational tests of the post-reionization IGM and the numerical simulations of reionization completed under the Cosmic Reionization On Computers (CROC) project. The CROC simulations match the gap distribution reasonably well, and also provide a good match for the distribution of peak heights, but there is a notable lack of wide peaks in the simulated spectra and the flux PDFs are poorly matched in the narrow redshift interval $5.5<z<5.7$, with the match at other redshifts being significantly better, albeit not exact. Both discrepancies are related: simulations show more opacity than the data.
\end{abstract}


\section{Introduction}
\label{sec:intro}

The Lyman-$\alpha$ forest, the collective intervening absorption systems from the spectra of distant quasars, has always been seen as an excellent probe of the intergalactic medium (IGM). Ever since its discovery, it has been instrumental in exploring the history of the IGM; the presence of spectral regions with low absorption has provided the first evidence for the existence of hydrogen reionization era \citep{croc:gunnpeterson}. After that discovery, constraining the epoch of reionization became an important topic in cosmological research \citep{croc:becker2015}.

A critical impact on the study of cosmic reionization with the Lyman-$\alpha$ forest has been made by the Sloan Digital Sky Survey (SDSS). A rapid rise in the flux decrement at $z\geq5.7$ in the spectra of quasars discovered by SDSS \citep{igm:bfws01,igm:fsbw06} has been the first and remains the strongest evidence that the epoch of reionization lies not far beyond that redshift mark.  Also, the distinct features of the $z\sim 6$ redshift QSO spectra - long, dark absorption gaps
separated by isolated transmission peaks - implies that reionization has largely ended at that point. More recent surveys of high-redshift QSOs \citep{croc:gallerani2008, croc:mortlock2011, croc:bolton2011, croc:banados2014, croc:mcgreer2015, igm:bbm15}, gammy ray bursts \citep{croc:chornock2013}, and high redshift galaxies \citep{croc:stark2010, croc:pentericci2011,croc:treu2013, croc:schenker2014,  croc:tilvi2014} strengthen this conclusion. 
Never-the-less, statistical properties of the Lyman-$\alpha$ forest in the very aftermath of cosmic reionization remain poorly explored. One of the reasons is that $z>5$ Lyman-$\alpha$ forest spectra look like spectra of no other astrophysical object; beginning students often take these absorption spectra for emission ones, confused by the narrow transmission regions between heavily blended absorption. As the result, many classical methods of spectral analysis (fitting absorption features, flux power spectrum, etc) become unusable or uninformative. New statistical probes of this unique cosmic phenomenon have to be developed, but some guidance on where to start is needed.

This is where cosmological simulations can come to the aid, complementing in many respects the direct observations of the high redshift IGM. Many of modern simulations give reasonable predictions for the overall statistical properties of the IGM (c.f.\ the column density or line width distributions, flux power spectrum, etc) at lower redshifts. Using simulations as testing grounds for exploring physical processes and to interpreting observational measurement is an important step in the scientific exploration. Of course, not any simulation is a suitable tool for interpreting the observational data - only those that reproduce the actual measurements well enough can be expected to provide an approximately correct physical model of reality. Hence, improving the simulation technology to be on par with the quality of observational data is an important precondition for any further progress.

In this paper we make a first small step in this direction, by comparing the statistical properties of the observed Ly-$\alpha$ forest with several sets of numerical simulations of reionization produced by the Cosmic Reionization On Computers (CROC) project \citep{ng:g14,ng:gk14}. Our goal is two-fold: first, to check how well the simulations match the general properties of the observed universe, as a validation of the simulation approach; second, to identify discrepancies and limitations of the simulations, from which one can, potentially, learn important insights into the underlying physics of reionization. 

This work should not be considered as a complete project or a self-contained effort. Provided the validation of the simulations against the observational constraints is acceptable (or, at the very least, the origin for any discrepancy is well understood), we can use simulations in the future as testing grounds for developing new statistical techniques and for exploring connections between observables and the properties of the underlying gas distribution.

\section{Simulations and Observational Samples}
\label{sec:simsanddata}

In this work we use simulations from the Cosmic Reionization On Computers (CROC) project \citep{ng:g14,ng:gk14}.  CROC simulations are designed to provide a physically reasonable and internally self-consistent model of reionization. They include a wide range of physical effects deemed necessary for modeling reionization correctly in cosmological hydrodynamic simulations, including star formation and stellar feedback, fully coupled 3D radiative transfer, non-equilibrium ionization of hydrogen and helium, radiation-field-dependent cooling and heating functions for the metal enriched cosmic gas, etc.

For this paper we use three independent realizations in $40h^{-1}\dim{Mpc}$ boxes from a new simulation series labeled ``Caiman''. These simulations are similar to previously published ones (such as, for example, runs B40.sf1.uv2.bw10.A-C from \citet{ng:g14}), but differ from them in one important respect - they utilize numerical convergence corrections as described in \citet{ng:g16a}, and, hence, offer a numerically converged model of reionization. In addition, we improved redshift sampling of the important time interval between $z=6.5$ and $z=5$ in the new runs, specifically to increase accuracy of modeling the post-reionization IGM.

CROC simulations from ``Caiman'' series achieve spatial resolution of $100\dim{pc}$ in physical units at all times. The initial conditions are set on the uniform grid of $1024^3$, resulting in the Nyquist wavenumber of $k_{\rm Ny} = 80h\dim{Mpc}^{-1}$ (corresponding to about $0.55\,({\rm km/s})^{-1}$ in velocity units at $z=6$), fully sufficient for modeling the Lyman-$\alpha$ forest even at lower redshifts.

These simulations account for the DC mode \citep{ng:gkr11} and, hence, properly model fluctuations up to the box size. However, the mean free path at the Lyman limit is about $(30-35)h^{-1}\dim{Mpc}$ at $z=6$ and $(55-60)h^{-1}\dim{Mpc}$ at $z=5$ \citep{igm:sc10}, comparable or exceeding the simulation box sizes. At present, we do not have simulations with larger box sizes at the redshifts of interest. Hence, with the current simulations, we cannot yet guarantee the lack of numerical artifacts due to the finite box size. In order to estimate the role of such artifacts, we also use a set of six realizations of a twice smaller, $20h^{-1}\dim{Mpc}$ box. Our science results, however, are all based on the $40h^{-1}\dim{Mpc}$ runs.
  
The simulation results presented here are largely a prediction (two of the three $40h^{-1}\dim{Mpc}$ simulations had been completed before the comparison with the data was made, and the third one was not in any way adjusted based on the preliminary comparison, but rather ran with the identical version of the code on an independent random realization of initial conditions). Never-the-less, some degree of tuning in the simulations is, indeed, present.

As has been shown in \citet{ng:g14}, the emission of ionizing radiation by model galaxies is controlled by a single parameter $\euv$, the ``escape fraction of ionizing radiation up to the simulation resolution''. The parameter $\euv$ encapsulates our ignorance about several important properties of real galaxies in the reionization era, such as the stellar IMF, the duration of the embedded stage of star formation, etc. Note, however, that since CROC simulations are reaching spatial resolution of $100\dim{pc}$, they are able to model the actual escape of ionizing radiation from the resolution scale to larger scales. The absorption of ionizing radiation from the stellar surface to the resolution limit (for example, absorptions inside a parent molecular cloud) is not resolved, and, hence, is also encapsulated into the $\euv$ parameter. This parameter cannot be deduced from the first principles yet, and has to be fixed by comparing with observational data.

The specific value of $\euv=0.15$ has been fixed by comparing the predictions for the mean Gunn-Peterson optical depth with the observational data from SDSS quasars \citep{igm:fsbw06} during the preparatory stage of CROC simulations, based on the results of $20h^{-1}\dim{Mpc}$ calibration runs, and before any production runs had been started. In appendix \ref{app:xhz} we show reionization histories for all three production $40h^{-1}\dim{Mpc}$ runs for the reference purposes.

A significant limitation of CROC simulations is the fidelity in modeling the quasar contribution. Since our simulation volumes are too small to sample the quasar population even approximately, quasars are included in CROC simulations only as a global background \citet[as explained in][]{ng:g14}. Hence, some of the effects from bright quasars, such as the change in the ionization state of the IGM inside their proximity zones, are not captured in the simulations.

In order to model the quasar absorption spectra, at each simulation snapshot we produce 1000 synthetic lines of sight. Each line of sight is $40h^{-1}\dim{Mpc}$ long (to make sure it fits entirely inside a $40h^{-1}\dim{Mpc}$ simulation box) and is oriented randomly with respect to coordinate axes. The latter is important for avoiding numerical biasing, which can arise if the lines of sight are parallel to principal directions of the box (since there are just 3 fundamental modes in a cubic box). Because of periodic boundary conditions, a simulation volume can be shifted by an arbitrary vector, so a randomly oriented line of sight starting at a random location can always fit entirely inside the simulation volume with the size equal to the line of sight length. 

In our smaller, $20h^{-1}\dim{Mpc}$ boxes, $40h^{-1}\dim{Mpc}$ long lines of sight do not fit entirely inside the simulation volume, and are continued beyond the box boundaries using periodic replicas of the simulation volume. Hence, the discrepancies between our $20h^{-1}\dim{Mpc}$ and $40h^{-1}\dim{Mpc}$ boxes can arise both from the insufficient simulation volumes and from the artifacts introduced in $20h^{-1}\dim{Mpc}$ runs due to lines of sight extending beyond the box boundaries. In appendix \ref{app:dc} we present tests that quantify some of these artifacts. We reemphasize here, only results from our $40h^{-1}\dim{Mpc}$ runs should be considered as ``science grade''.
 
For comparison with observations, we use two sets of data: flux probability distributions averaged over sightline segments (``skewers'') of comoving $40h^{-1}\dim{Mpc}$ - similar to those published by \citet{igm:bbm15}, but shorter to match the simulation box sizes, and the spectroscopic sample of SDSS quasars at $5.7 < z < 6.4$ from \citet{igm:fsbw06} and \citet{igm:bbm15}. The sample from \citet{igm:fsbw06} includes 19 luminous quasars with high S/N, moderate resolution spectroscopy. Twelve in the sample were observed using the Echellette Imaging Spectrograph (ESI) on Keck-II telescope. The ESI data have a spectral resolution between 2000 and 6000 (depending on the slit used during the observations). The rest of the spectra were observed on the MMT, HET and Kitt Peak 4-meter Telescopes, with spectral resolution of $R\sim 1000$. Details of this dataset can be found in \citet{igm:fsbw06}. 

To complement the data from \citet{igm:fsbw06}, we also use 7 quasar spectra from \citet{igm:bbm15} that were also obtained with ESI on Keck or X-Shooter on VLT with similar spectral resolution to the SDSS data.

Statistics of dark gaps and peak widths and heights is measured from these 26 spectra. Flux probability distributions in addition use observations of quasars J0841+2905, J0842+1218, J1621+5150 \citep{igm:jmf15,igm:jmf16}, SDSS0231-0728, SDSS0915+4244, SDSS1204-0021, SDSS1208+0010, and SDSS1659+2709 \citep{igm:bsr06,igm:bbh11,igm:bsr12}.

\section{Results}
\label{sec:res}

\subsection{Distribution of Optical Depth}
\label{sec:pdf}

\begin{figure}[t]
\includegraphics[width=\hsize]{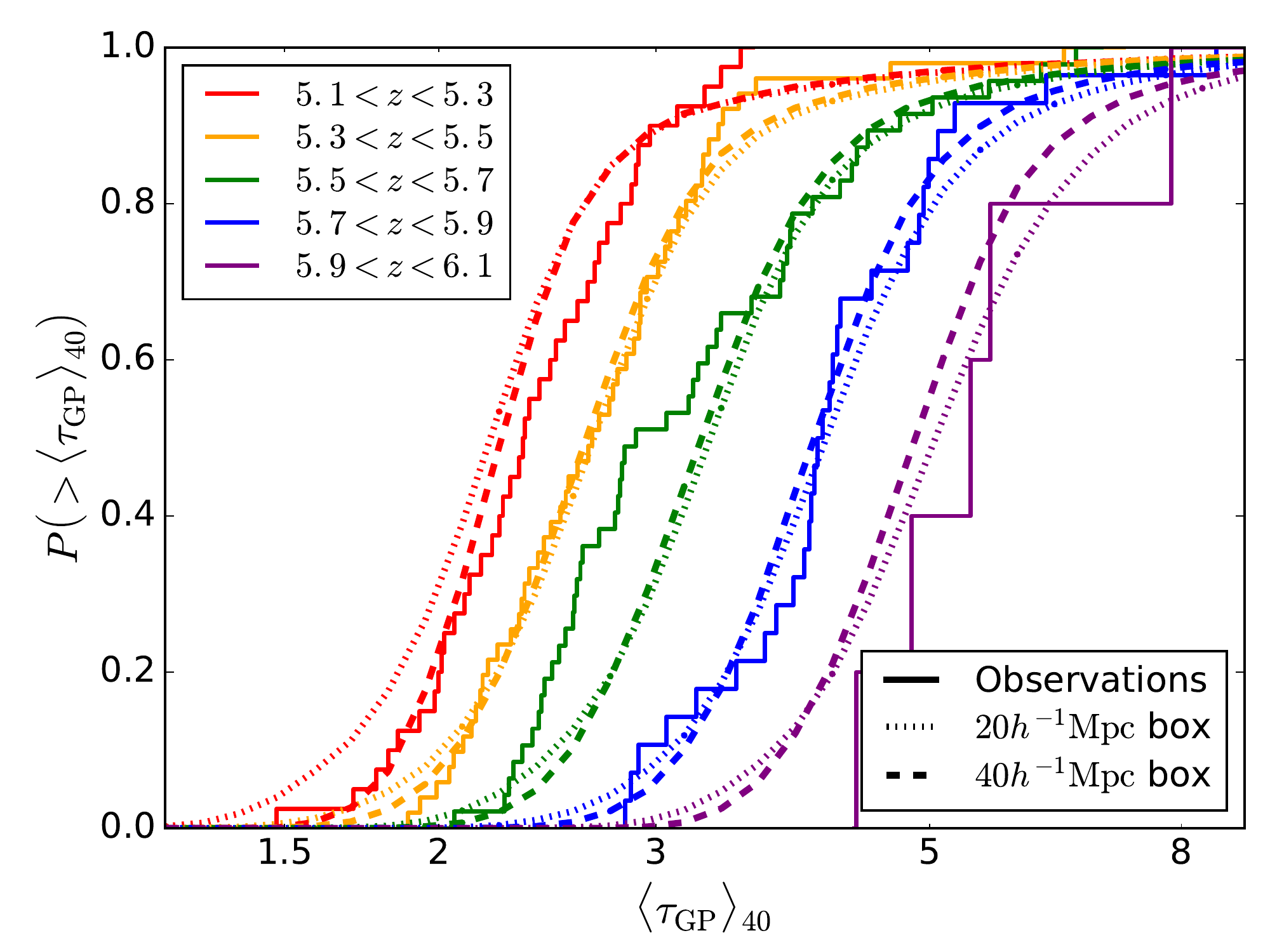}
\caption{Probability Distribution Functions for the effective IGM opacity in $40h^{-1}\dim{Mpc}$ skewers from our simulations (dotted and dashed lines) and observations of \protect\citet[][solid lines]{igm:bbm15}.\label{fig:poftau}}
\end{figure}

The simplest statistical description of any data is a one-point statistics, or the Probability Distribution Function (PDF). Since the transmitted flux along the quasar sightline is a stochastic field, the flux PDF depends on the smoothing scale for the spectrum. Thermal broadening provides the minimum possible smoothing scale for the transmitted flux, and it would be the ideal smoothing scale to use. Unfortunately, few observational data exist at such high spectral resolution. 

Recently, \citet{igm:bbm15} published the flux PDF averaged over sightline segments (``skewers'') of comoving $50h^{-1}\dim{Mpc}$ in length in several redshift bins all the way to $z\approx6$. While it is a less detailed probe of the post-reionization IGM than a PDF of a fully resolved spectrum, the distance of comoving $50h^{-1}\dim{Mpc}$ is comparable to the mean free path for an ionizing photon at $z<6$ \citep{igm:sc10}, and, hence, is a sensitive probe of the fluctuations in the ionization state of the IGM, as has been demonstrated in \citet{igm:bbm15}.

In this work we use a similar statistic, except that we reduce the skewer length to $40h^{-1}\dim{Mpc}$ to fit entirely into our simulation boxes, as explained in the previous section. Figure \ref{fig:poftau} shows the comparison of CROC simulations to the observational measurement in four redshift bins. We show both $20h^{-1}\dim{Mpc}$ and $40h^{-1}\dim{Mpc}$ simulation sets in Figure \ref{fig:poftau} to demonstrate the degree of (or lack thereof) convergence of the numerical results. 

It appears that as redshift decreases, the agreement between the simulations and the data is becoming progressively worse, although a better agreement at higher redshift may be artificial and simply due to larger sparsity of the observational data. Another apparent feature is a large discrepancy for $5.5<z<5.7$ at low optical depths. While simulations results seem to ``march'' to higher optical depths with increasing redshift more-or-less uniformly, the data show a different trend: higher optical depths regions follow simulations, while low optical depths segments seem to ``lag behind''. We discuss this important discrepancy in more detail in the conclusions.

\subsection{Statistics of Dark Gaps}
\label{sec:gaps}

Going beyond PDFs, two-point statistics (power spectrum, correlation function) often used in studies of the large scale structure and the Lyman-$\alpha$ forest at intermediate redshifts ($z\sim2-4$). The transmitted flux in the spectra of high redshift quasars ($z\ga5$) is so far from a Gaussian random field, however, that these statistics include only a small amount of information about the distribution of neutral gas in the universe. Hence, traditionally other, more sophisticated statistical probes have been used to explore the forest in the very aftermath of cosmic reionization. 

One such probe is the distribution of dark gaps (continuous regions of low flux). It has been measured observationally \citep{igm:fsbw06,igm:gcf06} and has also been used as an effective indicator for comparison of observations and simulations \citep{croc:paschos2005,igm:gff08}. It provides a natural description of statistical properties of high redshift Lyman-$\alpha$ spectra (which mostly contain single transmission spikes separated by dark gaps of various lengths) and shows well the drastic evolution in absorption between $z\approx5.5$ and $z\approx6$.

Following the definition of dark gaps from \citet{croc:song2002} and \citet{igm:fsbw06}, we analyze the distribution of dark gaps in our simulation sets and compare it to the sample of 12 Sloan Digital Sky Survey (SDSS) quasars from \citet{igm:fsbw06}. We select only quasars obtained with the Keck ESI instrument to have uniform spectral sampling and noise properties in the data sample.

The observational sample includes 86 gaps in the interval $5.3<z<5.5$, 77 gaps in the interval $5.5<z<5.7$, 46 gaps in the interval $5.7<z<5.9$ and 22 gaps in the interval $5.9<z<6.1$. Due to spectral binning with $R=2000$, only gaps larger than about $2\dim{Mpc}$ are detectable both in the data and in the simulations.

\begin{figure*}[t]
\includegraphics[width=0.5\hsize]{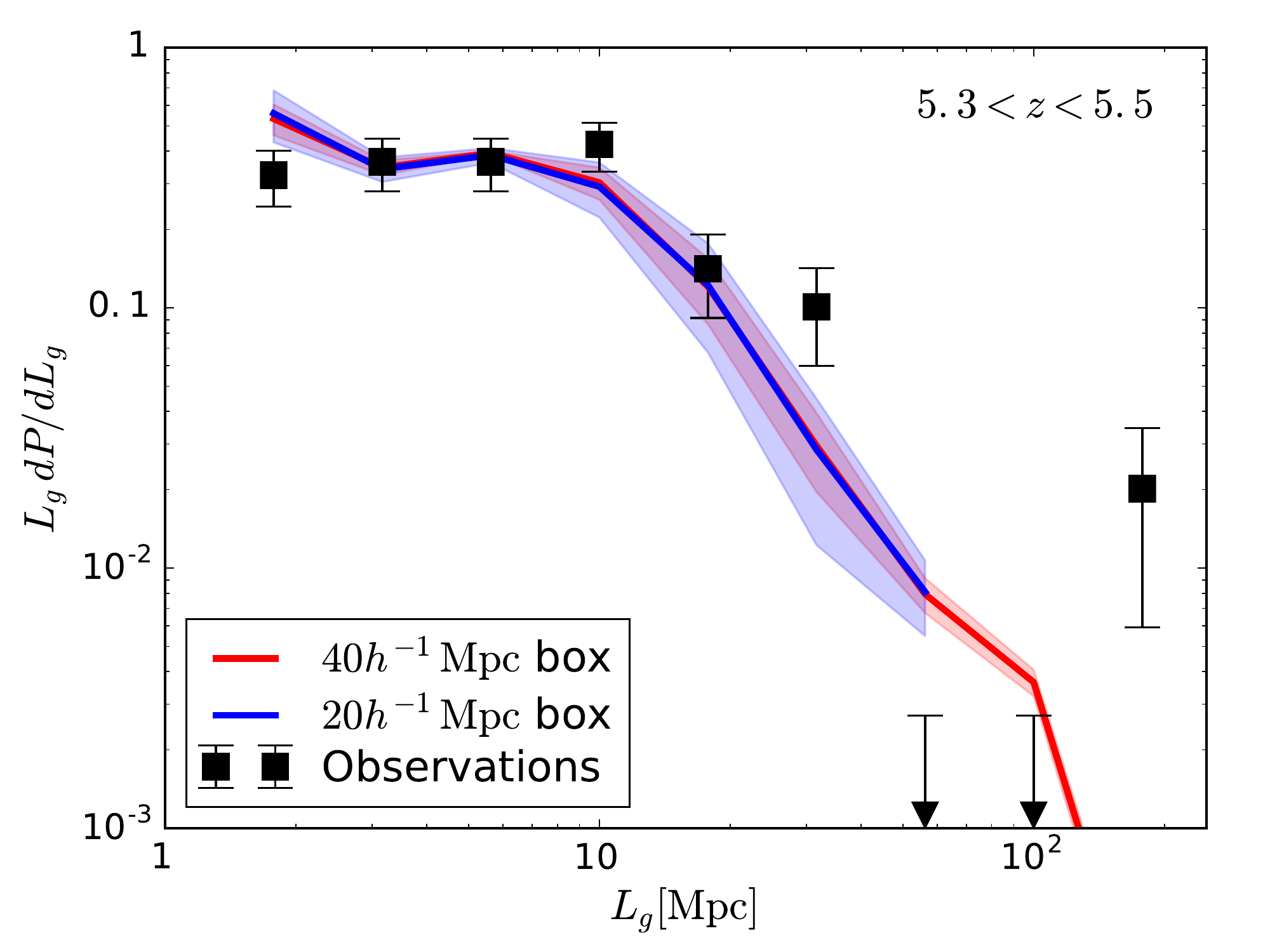}%
\includegraphics[width=0.5\hsize]{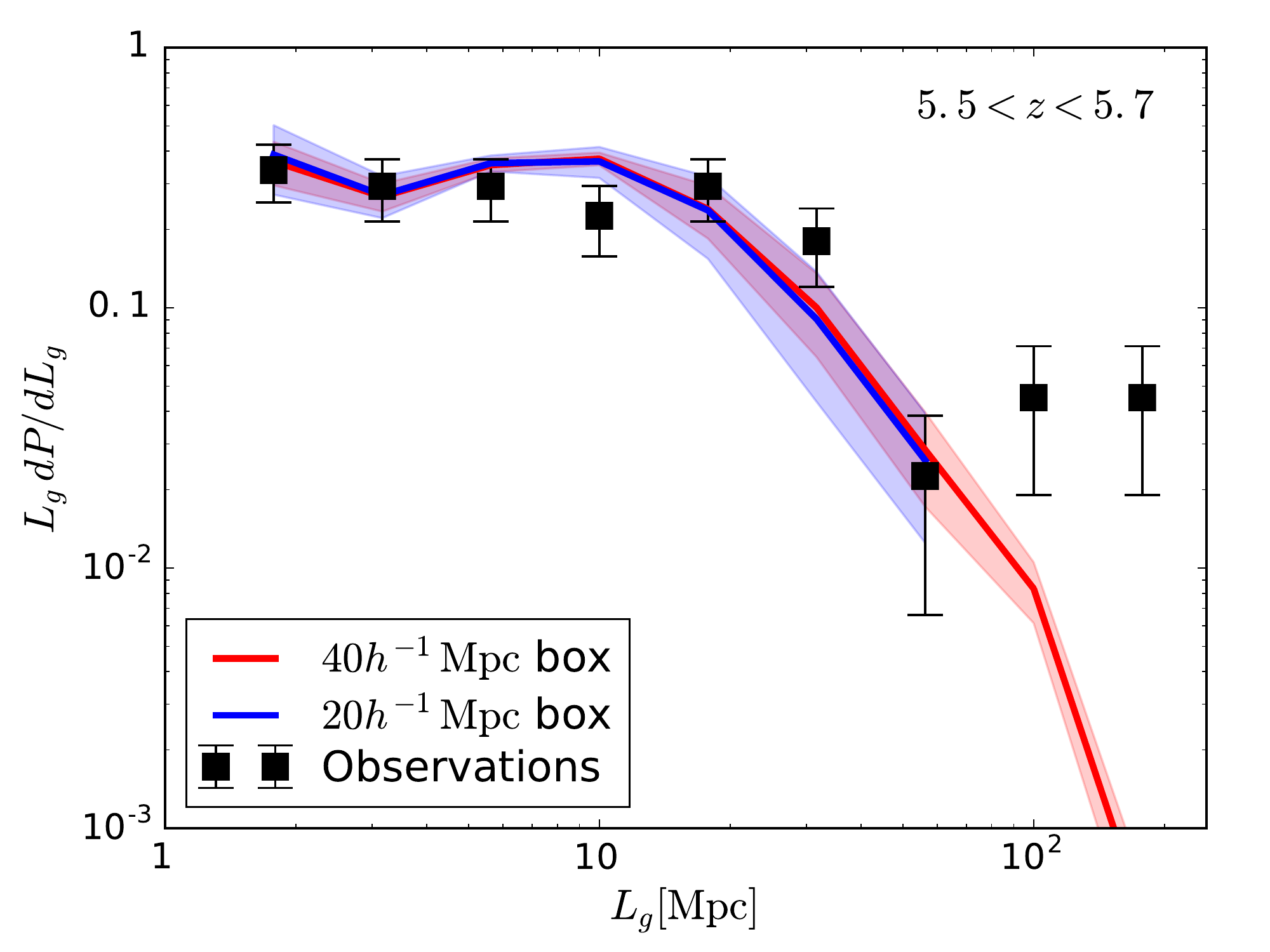}\newline
\includegraphics[width=0.5\hsize]{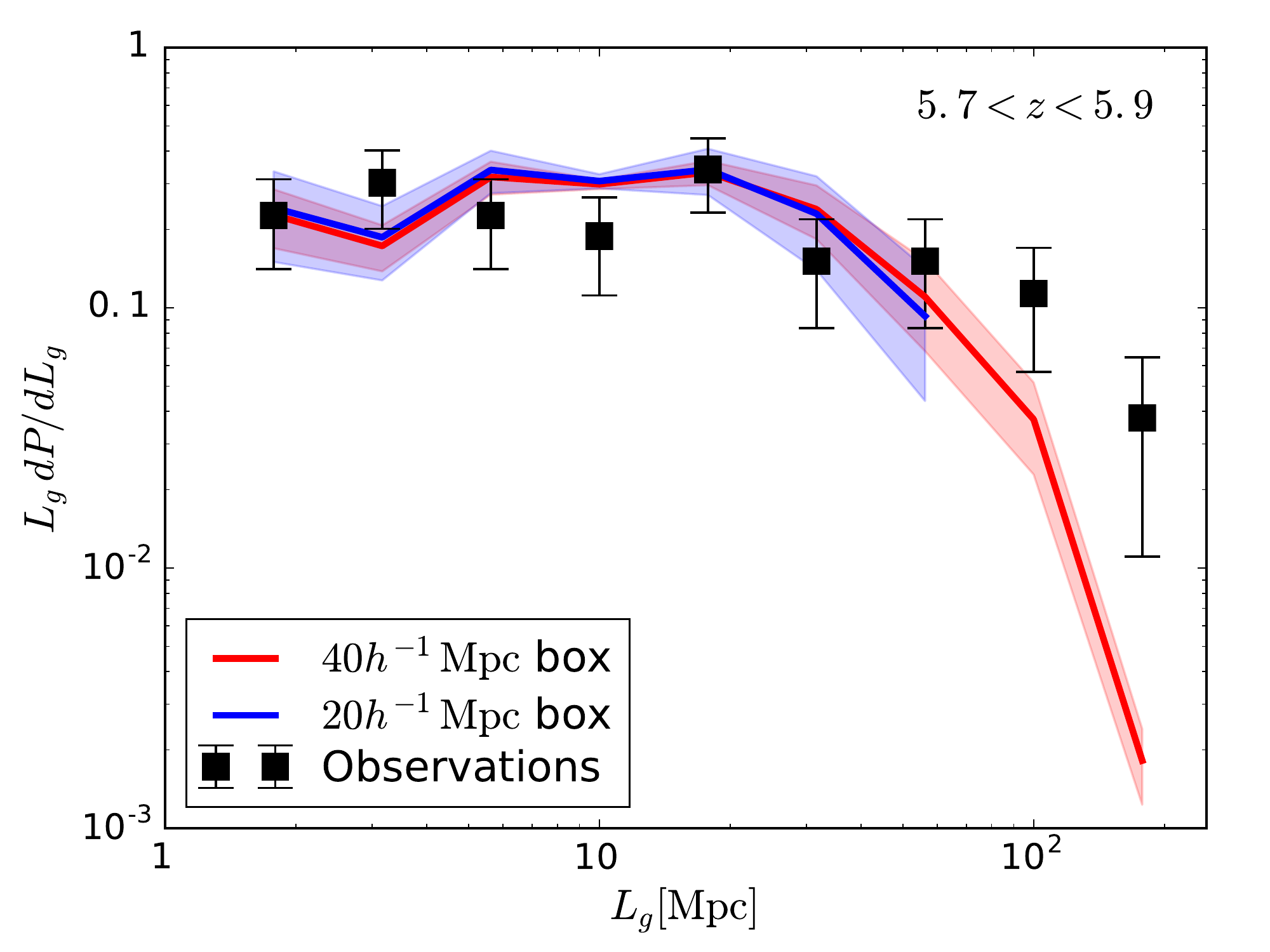}%
\includegraphics[width=0.5\hsize]{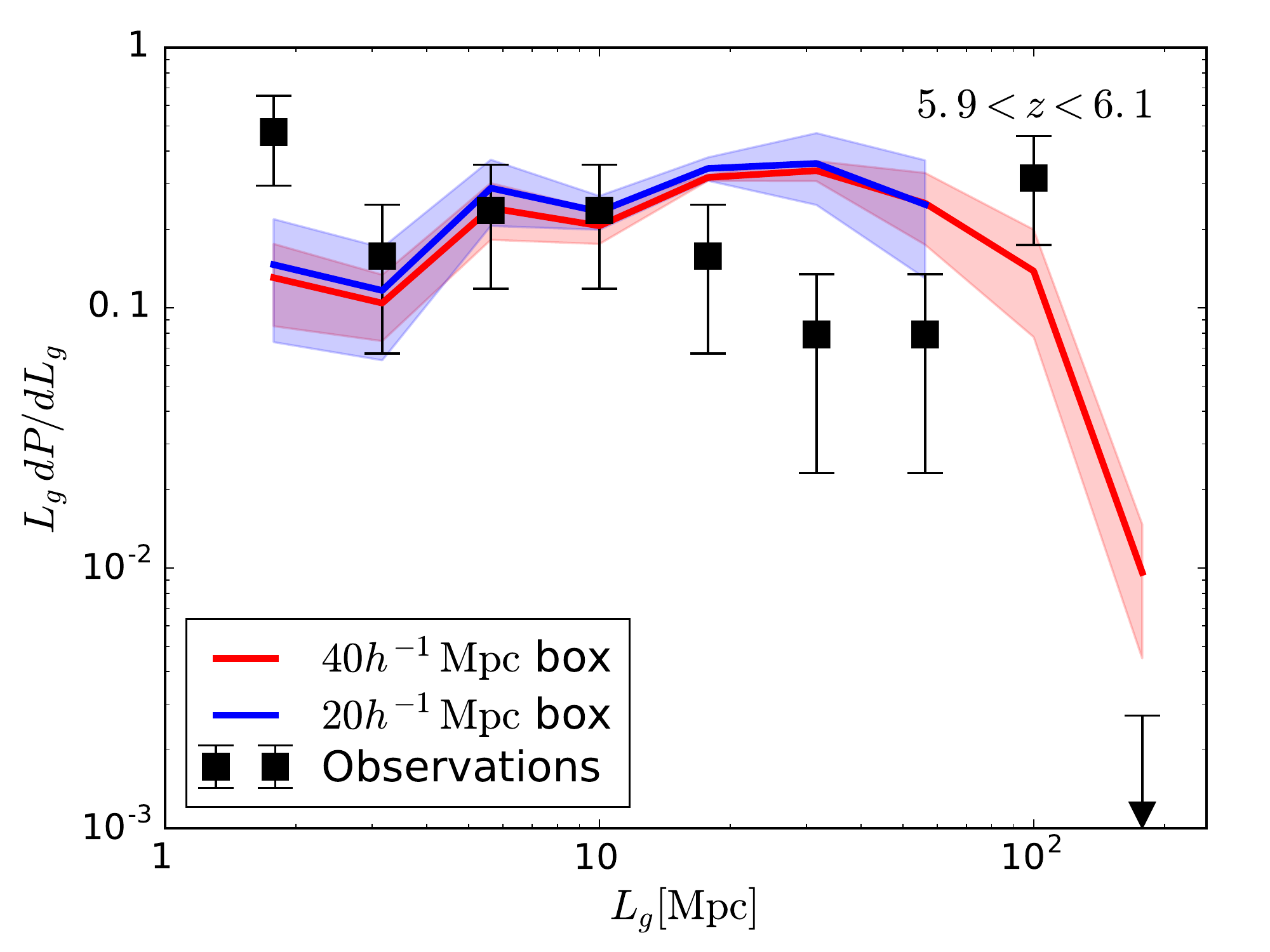}%
\caption{Distribution of dark gaps for two simulation sets with different box sizes and the observational data from SDSS. Shaded bands show the error of the mean for the simulated results.\label{fig:gap_4} }
\end{figure*}

In order to make the comparison between the observations and the simulations more realistic, we add noise to synthetic spectra sampled directly from the observational measurements. Specifically, at each simulation snapshot we take the rms noise in all observed spectra that cover the snapshot redshift in the redshift interval $\Delta z = \pm 0.1$ around the snapshot epoch and add noise values from different observed sightlines in quadratures. We then add Gaussian noise with the observed rms to all synthetic spectra.

Observed spectra come with slightly variable spectral resolution at different wavelength. To make spectral resolution between simulations and data exactly the same, we bin both real and synthetic spectra in wavelength bins of uniform resolution $R\equiv\lambda/\Delta\lambda=2000$, which is approximately the lowest common resolution of all observed spectra. The reason for using binning rather than, say, Gaussian filtering, is that binning is independent of any prior finer binning (i.e.\ binning a $R=4000$ spectrum with $R=2000$ resolution results in the $R=2000$ spectrum), while in Gaussian smoothing window widths add in quadratures, and hence any prior smoothing needs to be known precisely.

Spectral gap statistics uses a single parameter - the threshold value that specifies what ``low flux'' means. This value is commonly quantified by the effective optical depth $\tau_{\rm min}$ - a contiguous region in the spectrum with flux $F < \exp(-\tau_{\rm min})$ is considered to be a single dark gap. We use the value of $\tau_{\rm min}=2.5$ as the fiducial value, to be consistent with \citet{igm:fsbw06}.

Figure \ref{fig:gap_4} shows the comparison of dark gaps in four different redshift ranges between our simulations and the SDSS data (we do not show the $5.1<z<5.3$ redshift range that is not covered by the quasars in the observational sample). In order to make the comparison between a few long lines of sight from observations data and a large number of relatively short lines of sight from the simulations meaningful, we plot the differential probability $L_g d/dL_gP$ to find a dark gap of a given comoving length between $L_g$ and $L_g+dL_g$ along a given sightline. Operationally, for a given number of sightlines, we find the number $\Delta N(L_g,\Delta L_g)$ of gaps that fall in a bin of length $\Delta L_g$ around $L_g$, and define
\[
  \frac{dP}{dL_g} = \frac{\Delta N(L_g,\Delta L_g)}{N_{\rm tot}\Delta L_g},
\]
where $N_{\rm tot}$ is the total number of all gaps (of any length), so that $dP/dL_g$ is normalized to unity,
\[
  \int \frac{dP}{dL_g}\, dL_g = 1.
\]
With such definition, $dP/dL_g$ is independent of the bin size $\Delta L_{\rm gap}$ in the limit of many sightlines and small bins. 

Lengths of dark gaps in the simulations are limited by the finite sizes of computational boxes, and that restricts the dynamic range of curves plotted in Figure \ref{fig:gap_4}, which is especially apparent for the smaller, $20h^{-1}\dim{Mpc}$ boxes. Overall, the agreement between simulations and observations is good in all four redshift ranges, although it appears to get slightly worse in the lowest redshift bin. The apparent discrepancy can be caused by small number statistics (the last bin in the top left panel includes just 1 gap), but also can be genuine, consistent with the limitation of CROC simulations discussed below.

\begin{figure}[t]
\includegraphics[width=\hsize]{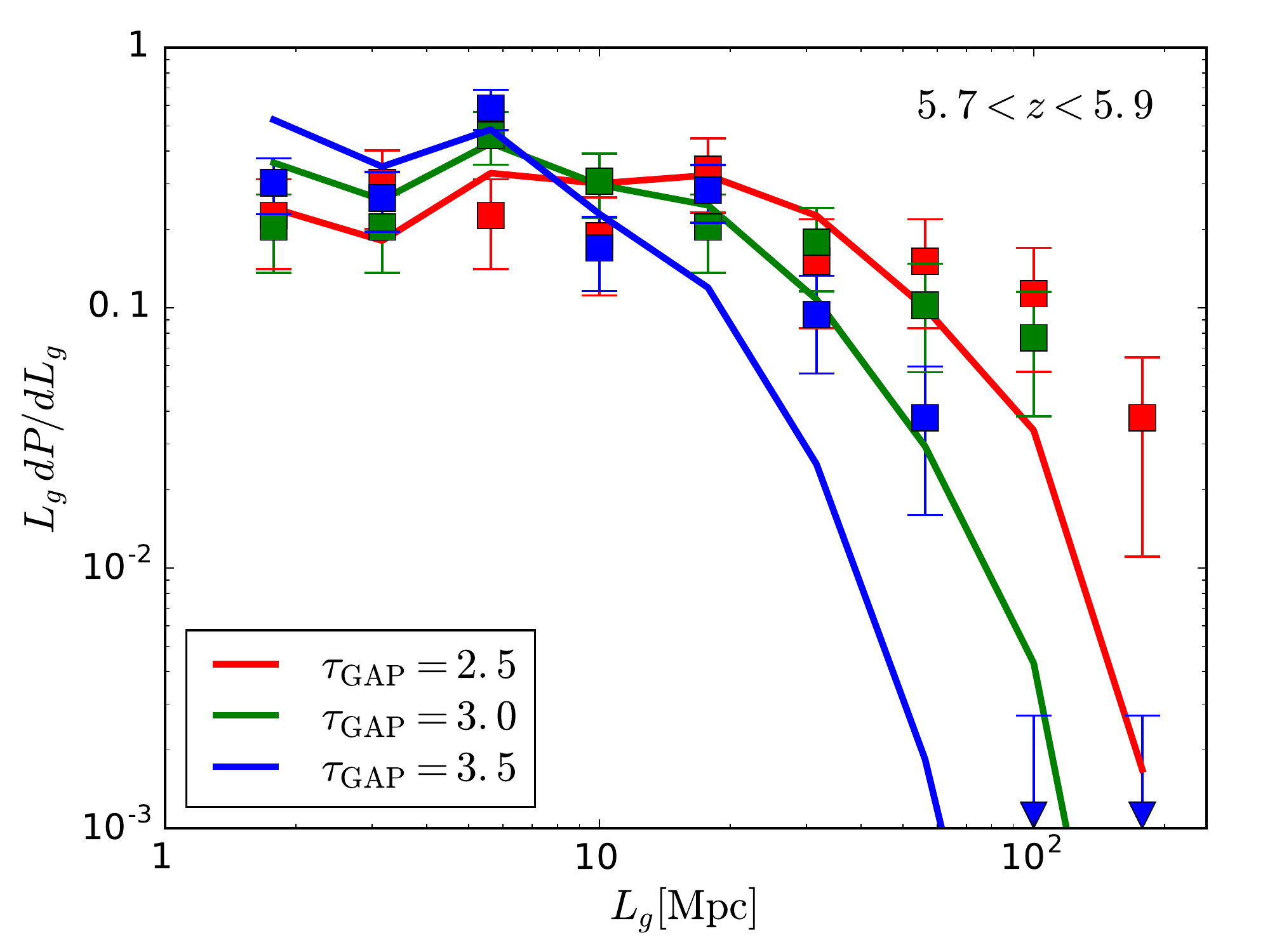}%
\caption{Distribution of dark gaps in the redshift interval $5.5<z<5.7$ for three different values of the threshold $\tau_{\rm min}$.\label{fig:gap_threshold}}
\end{figure}

We also explore the effect of varying the dark gap threshold $\tau_{\rm min}$ on the overall gap distribution in Figure \ref{fig:gap_threshold}. As could be expected, as the flux threshold $\exp(-\tau_{\rm min})$ is being lowered, longer gaps split into shorter ones, shrinking the tail of the distribution toward lower values. However, the overall shape of the distribution does not change that much.

\subsection{Statistics of Transmission Peaks}
\label{sec:peaks}

While dark gap distribution is a powerful tool that has been successfully used to characterize the post-reionization IGM, it has its limitation. First, it provides little information on the density distribution in the IGM, as most densities except the lowest ones would produce dark gaps at $z\sim6$. Second, transmission peaks that separate gaps are treated just as walls, with no significance given to the peak height (besides it is being above the threshold) or its width. Hence, the distribution of transmission peak heights and widths may serve as a complementary statistics.

In order to explore this new statistical test of the post-reionization IGM, we define as a ``peak'' a continuous segment of the spectrum, such that for all pixels in the segment the normalized flux is above $\alpha h_p$, where $h_p$ is the maximum normalized flux value inside the segment. The width of the segment $w_p$ is then identified with the ``width of the peak''. The value of $\alpha$ could be signal-to-noise dependent, but in this pilot study we choose a fixed threshold of $\alpha=0.5$, so that the peak width is the familiar Full Width at Half Maximum (FWHM). That particular choice is entirely empirical, and we find that within the reasonable range  (if $\alpha$ is not too small), the threshold value $\alpha$ affects the peak height and width statistics rather little (besides a trivial rescaling of widths). To avoid including noise, we only consider a peak to be real if it is wider than one resolution element and has a signal-to-noise ratio of at least 3.

\begin{figure}[t]
\includegraphics[width=\hsize]{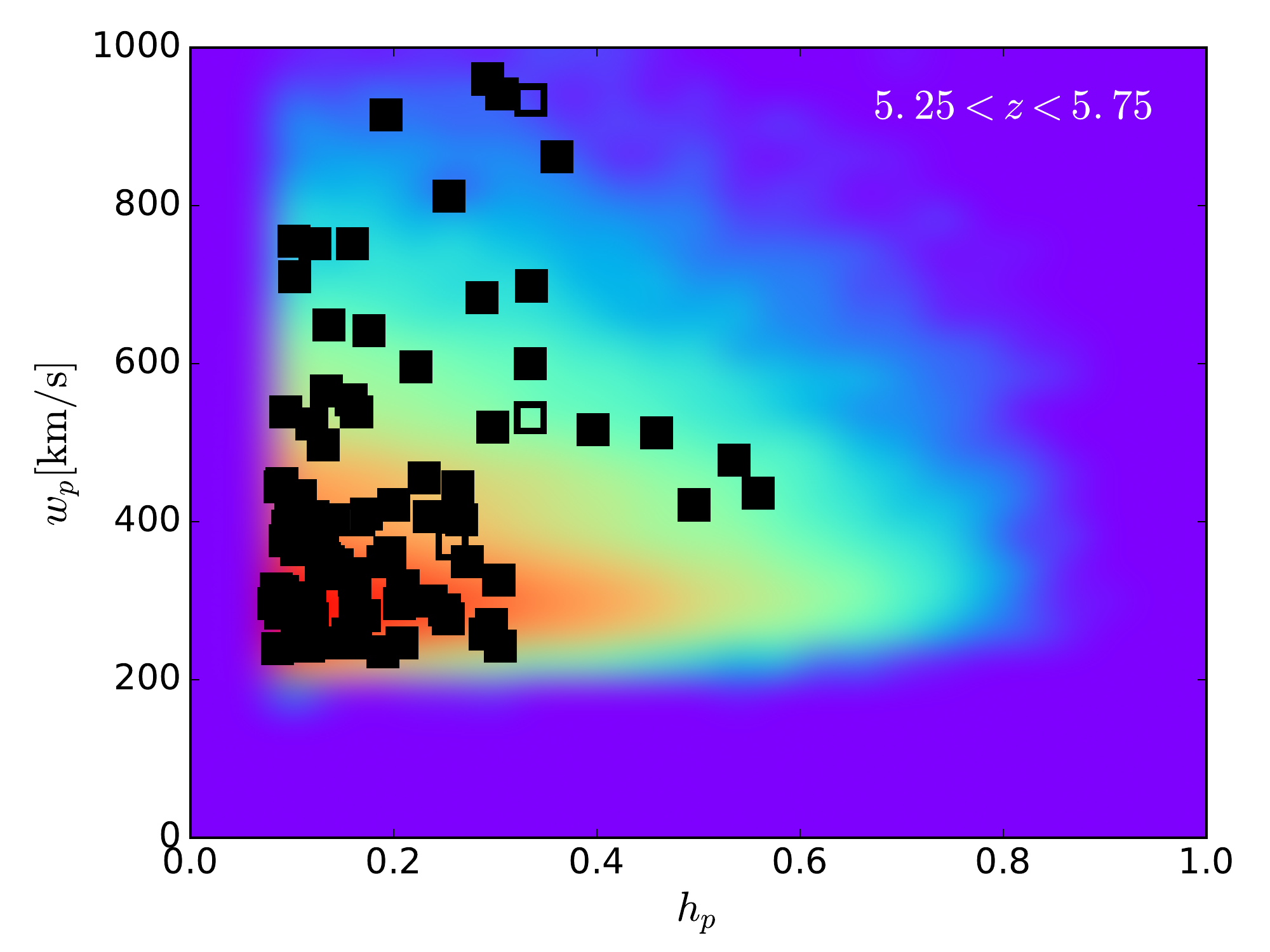}\newline
\includegraphics[width=\hsize]{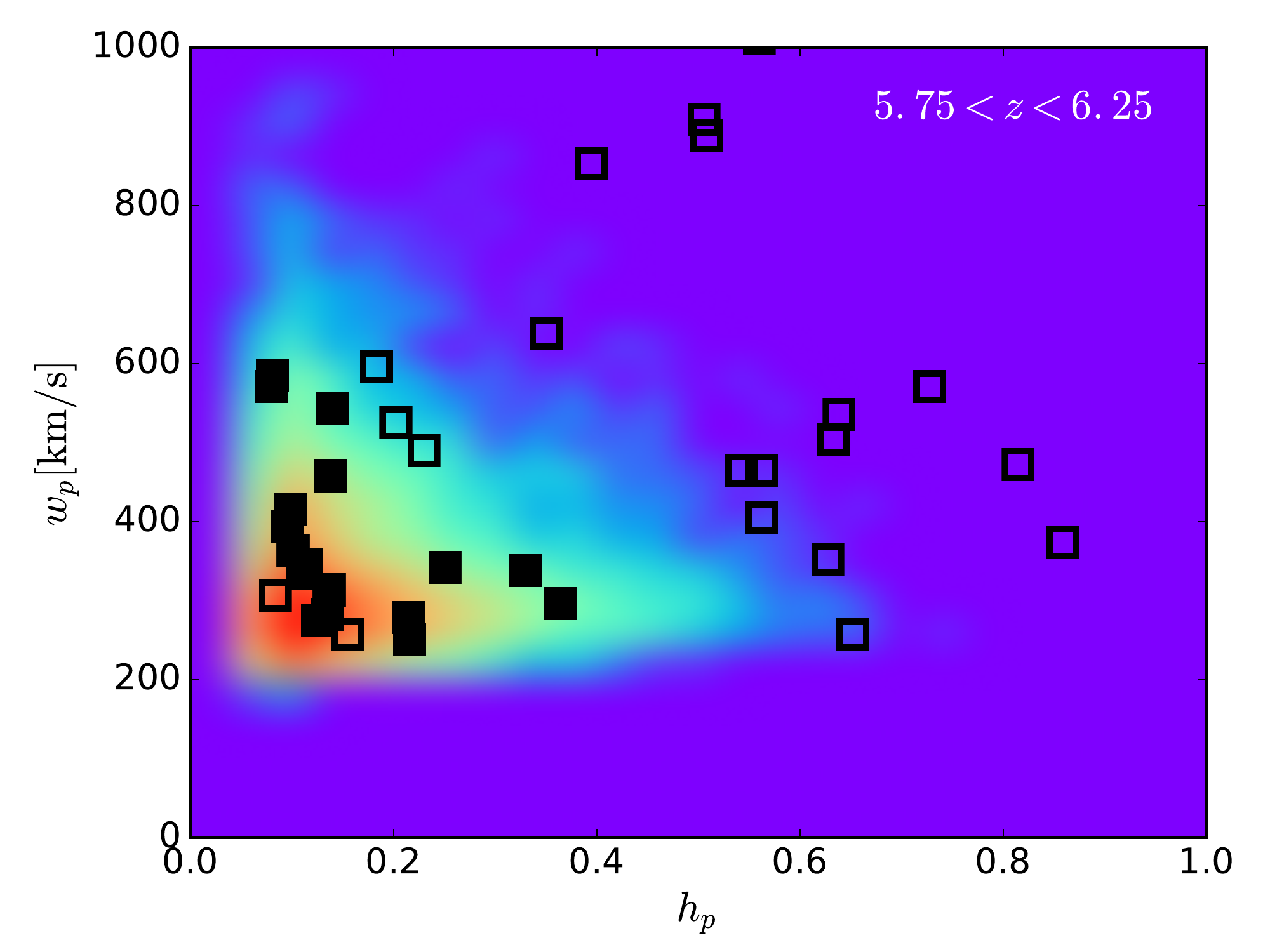}%
\caption{Two-dimensional distributions of transmission peaks and widths in the simulations (colored images) and the data (squares). Filed squares show observed peaks outside the proximity zones of their quasars, while open squares are for peaks inside the quasar proximity zones.\label{fig:peak2d}}
\end{figure}

In Figure \ref{fig:peak2d} we present the peak height and width distributions as 2D histograms, both for the SDSS observational data and for our $40h^{-1}\dim{Mpc}$ simulation set, in two redshift intervals. Since quasars affect their own environments within their ``proximity zones'', the ionization state of the gas near quasars is expected to be different from the general IGM. In order to check for that effect, we show in Fig.\ \ref{fig:peak2d} peaks within and outside the quasar proximity zones with different symbols. Sizes of proximity zones for for the sample we study in this paper were measured by \citet{rei:cwf10}; we use their values for all quasars in the sample except for J1306+0356, where visual examination of the data indicates that the proximity zone extends to $10\dim{Mpc}$, beyond the value of $6\dim{Mpc}$ quoted by \citet{rei:cwf10}. The proximity effect is glaring at $z\sim6$, but is much less pronounced at $z\sim5.5$. 

\begin{figure*}[t]
\includegraphics[width=0.5\hsize]{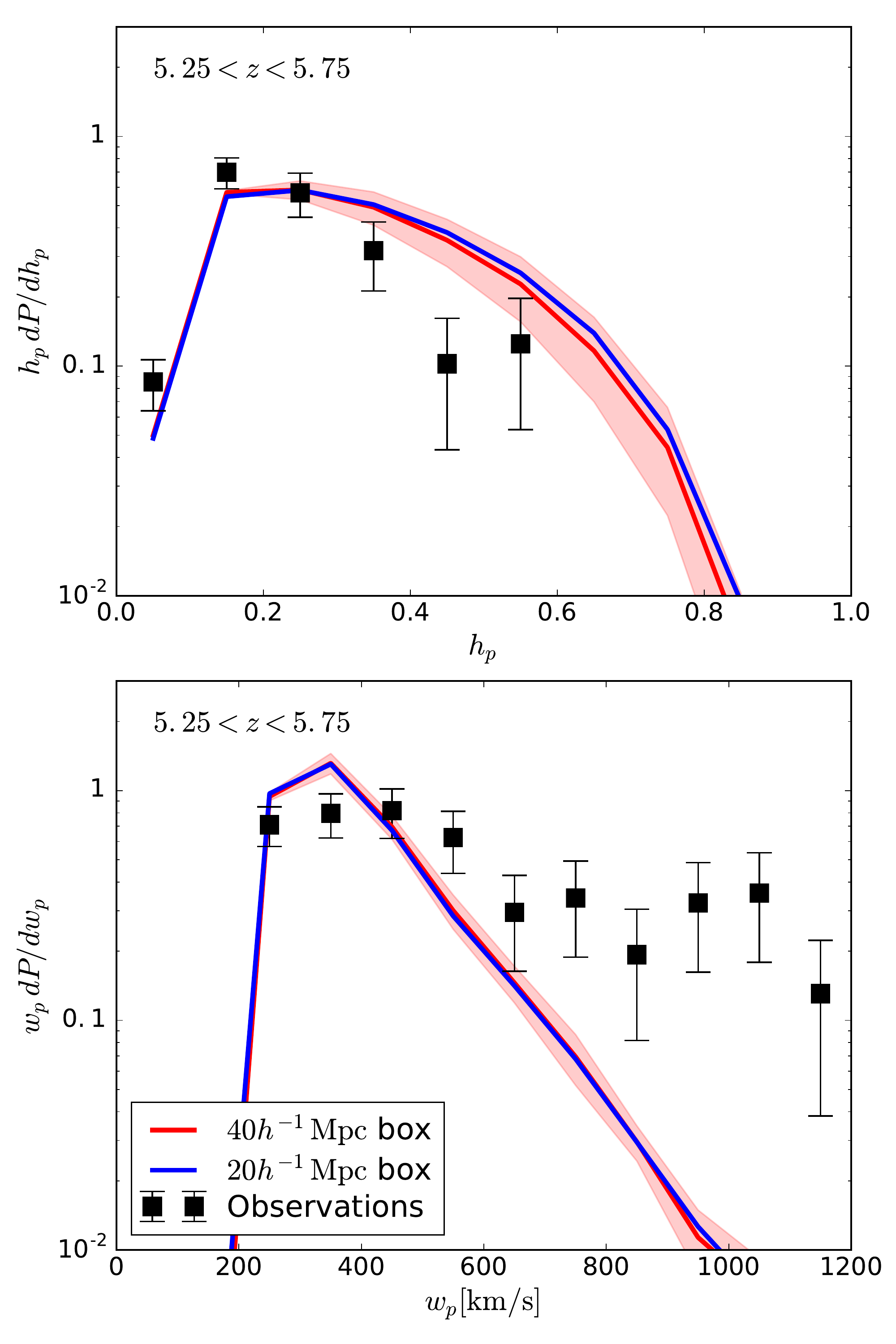}%
\includegraphics[width=0.5\hsize]{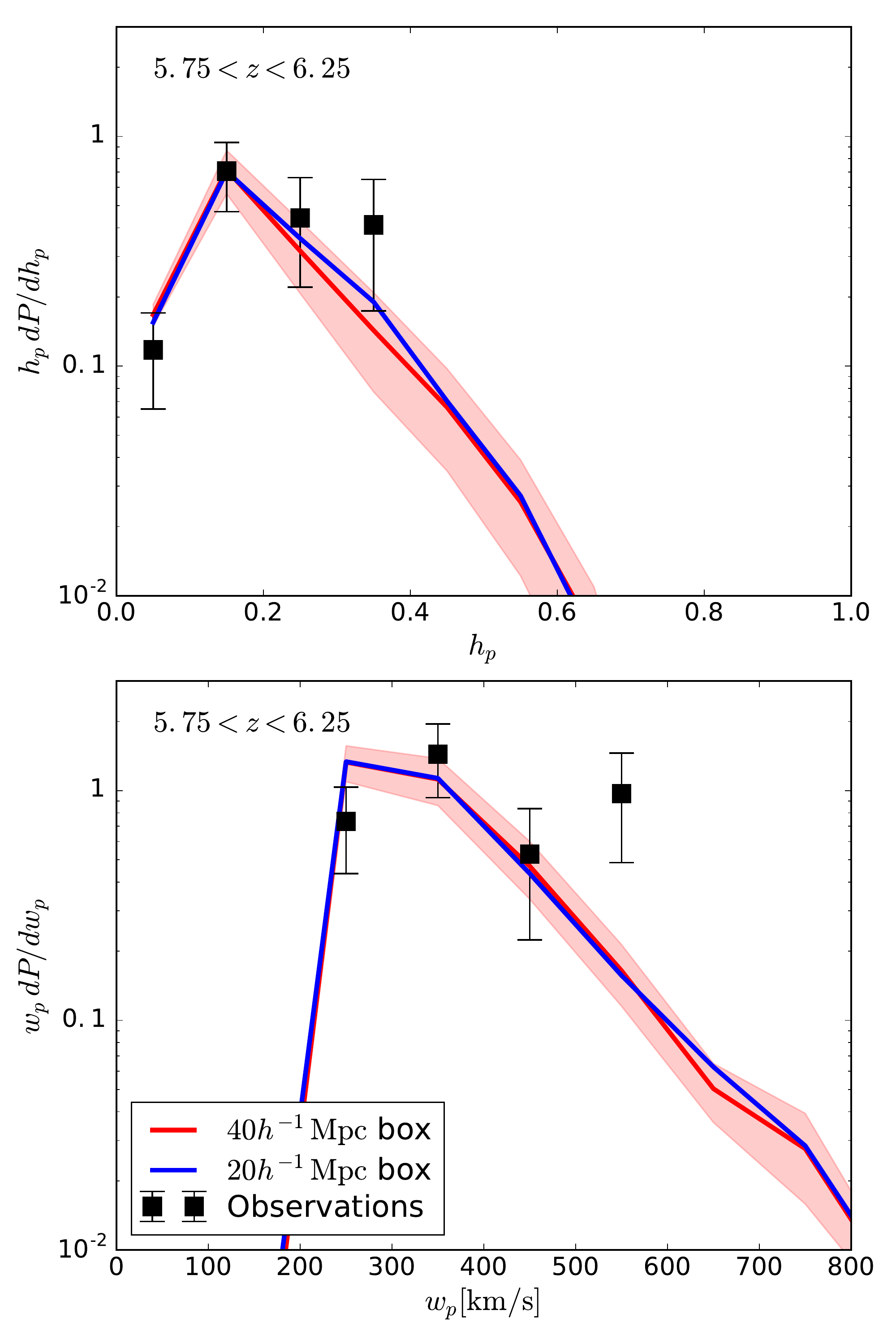}%
\caption{One-dimensional probability distributions for peak heights and widths in the two redshift bins in two simulation sets and in the data. Shaded bands show the error of the mean for the simulated results. The discrepancy at large $w_p$ is real and is not caused by the limited sizes of the simulation volumes.}
  \label{fig:pdfs}
\end{figure*}

A more quantitative comparison of these distributions is offered in Figure \ref{fig:pdfs}, which shows one dimensional (i.e.\ projected along one of the two axes) probability distributions for peak heights and widths. While the peak heights are reproduced faithfully (within the observational uncertainties) by the simulations, there is a significant lack of wide peaks in the synthetic spectra. This discrepancy cannot be attributed to the limited volumes of CROC simulations - for both distributions the differences in the simulation volume are much smaller than the magnitude of the discrepancy. We also note that observed spectra have spectral resolution better than $R=2000$, so transmission peaks with $w_p>800\dim{km/s}$ are well resolved; hence, the discrepancy we find cannot be attributed to instrumental effects and is genuine.

\section{Conclusions}
\label{sec:concl}

Statistical properties of the Lyman-$\alpha$ forest in high redshift ($z\ga5$) quasar spectra is a barely touched, full of unexplored potential research subject. Spatial variations of the transmitted flux in the spectra contain rich information on the physics of the IGM and the cosmological density distribution, as well as on the large-scale clustering properties of reionizing sources.  In this exploratory work we have made a modest attempt at accessing this information, and compared CROC simulations with some of the existing the observational data on high redshift quasars. 

Besides applying the commonly used statistical tests: the PDF for the large-scale opacity (most sensitive to the large-scale distribution of ionizing sources) and the dark gap distribution (most sensitive to the large-scale matter clustering), we also use additional statistical probes of the small-scale structure in the IGM: heights and widths of transmission peaks \citep{igm:gff08}.

The CROC simulations match the gap distribution reasonably well, and also provide a good match for the distribution of peak heights, but there is a notable lack of wide peaks in the simulated spectra and the flux PDFs are poorly matched in the narrow redshift interval $5.5<z<5.7$, with the match at other redshifts being significantly better, albeit not exact. Both discrepancies are related: simulations show more opacity than the data.

The source of this discrepancy remains unclear at present. One possible reason is proximity zones around foreground quasars, as has been previously proposed by \citet{igm:gff08}. While we exclude all transmission peaks within the proximity zones of targeted quasars, a sightline can also traverse a proximity zone around a foreground quasar, which would create a broad transmission spike. As is mentioned above, in CROC simulations quasars are included only as a global background, and, hence, individual proximity zones are not modeled. Another potential source of discrepancy is a limited simulated volume. While Fig.\ \ref{fig:pdfs} shows that the dependence on the simulation box size is less than the discrepancy with the data, with only two sampled box sizes the concordance may be accidental and simulations may, in fact, underpredict the abundance of large peaks. Exploring this discrepancy fully will require a substantial amount of effort for an uncertain benefit (surely, no one is going to claim that our simulations present the final truth, thus, discrepancies with the data are expected), and, therefore, we do not proceed with that effort at present.

An interesting avenue of further research is to use numerical simulations to understand better what physical properties of the IGM control the shapes of transmission peaks. For example, one may discover in the future that transmission peaks of particular heights and width are especially sensitive to some physical parameters, like gas density, ionization state, temperature, or the local ionization rate. Only in that case an effort for better modeling the peaks of those particular properties will be warranted.

\acknowledgements 
Fermilab is operated by Fermi Research Alliance, LLC, under Contract No.~DE-AC02-07CH11359 with the United States Department of Energy. This work was also supported in part by the NSF grant AST-1211190 and by the Munich Institute for Astro- and Particle Physics (MIAPP) of the DFG cluster of excellence "Origin and Structure of the Universe". Simulations used in this work have been performed on the National Energy Research Supercomputing Center (NERSC) supercomputers ``Hopper'' and ``Edison'' and on the Argonne Leadership Computing Facility supercomputer ``Mira''. An award of computer time was provided by the Innovative and Novel Computational Impact on Theory and Experiment (INCITE) program. This research used resources of the Argonne Leadership Computing Facility, which is a DOE Office of Science User Facility supported under Contract DE-AC02-06CH11357. This work made extensive use of the NASA Astrophysics Data System and {\tt arXiv.org} preprint server.

\appendix

\section{Reionization Histories of Production Runs}
\label{app:xhz}

\begin{figure}[t]
\includegraphics[width=0.5\hsize]{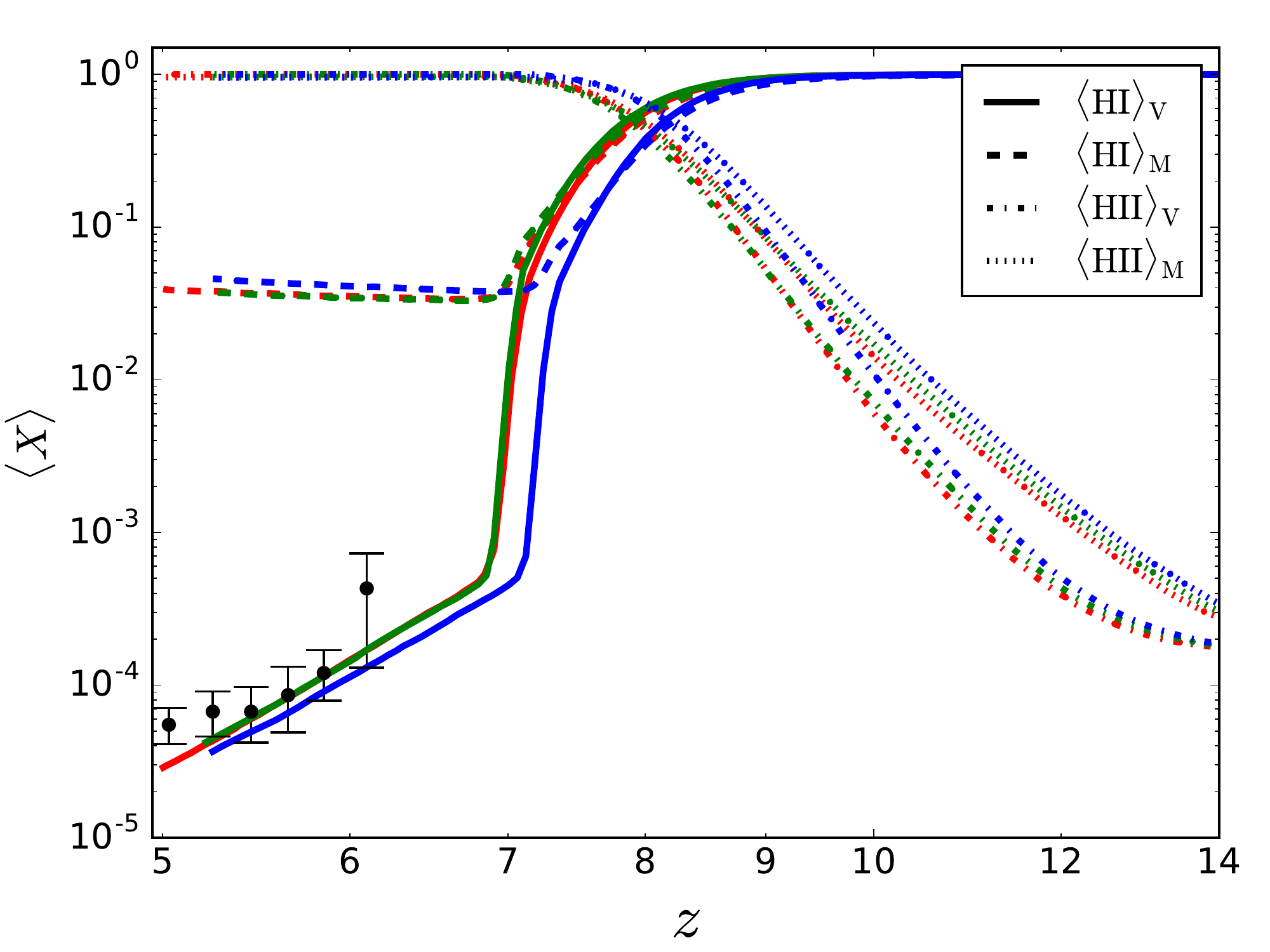}%
\includegraphics[width=0.5\hsize]{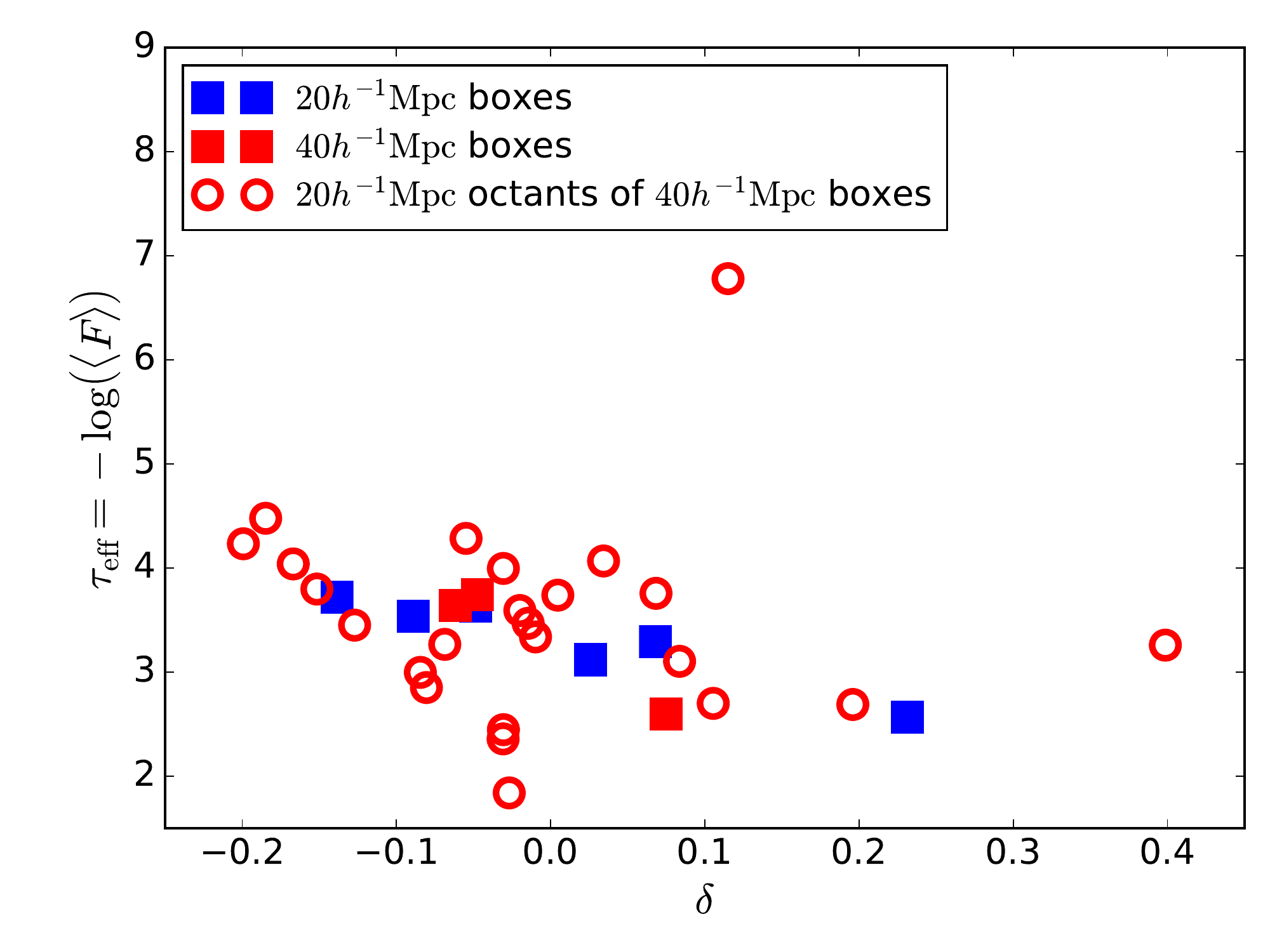}
\caption{Left: ionization histories for three $40h^{-1}\dim{Mpc}$ production runs (shown in different colors). Right: average Ly-$\alpha$ opacity vs mean overdensity in six $20h^{-1}\dim{Mpc}$ boxes (blue squares), three $40h^{-1}\dim{Mpc}$ boxes (red squares), and 24 $20h^{-1}\dim{Mpc}$-sized octants of three $40h^{-1}\dim{Mpc}$ boxes (red open circles) at $z=5.7$.}
  \label{fig:app}
\end{figure}

The focus of this paper is Ly-$\alpha$ forest in the post-reionization universe. However, knowing the full reionization histories may be useful in interpreting the difference between the individual simulations. In the left panel of Figure \ref{fig:app} we show mass- and volume-weighted reionization histories for all three production runs used in this work, just as a reference.

\section{Test of Numerical Artifacts due to DC mode}
\label{app:dc}

As we discussed above, our simulations boxes are only large enough to be comparable to the mean free path for ionizing radiation, and, hence, may introduce numerical artifacts. Following the referee's suggestion, we perform here a test that aims at checking the role of the DC mode in modulating the observed signal.

We do not yet have large volume to check our $40h^{-1}\dim{Mpc}$ boxes, but we can test whether our individual $20h^{-1}\dim{Mpc}$ boxes serve as a fair representation of $20h^{-1}\dim{Mpc}$ - sized octants of $40h^{-1}\dim{Mpc}$ boxes. To that end, we compute the average Ly-$\alpha$ opacity for each of our six $20h^{-1}\dim{Mpc}$ (averaged over 1000 $20h^{-1}\dim{Mpc}$ skewers) and in each of 8 octants for both $40h^{-1}\dim{Mpc}$ boxes (also averaged over 1000 $20h^{-1}\dim{Mpc}$ skewers) and compare the average Ly-$\alpha$ opacities with mean overdensity $\delta$ in each box or octant.

The result of such comparison is shown in the right panel of Figure \ref{fig:app}. Red open circles show 24 octants from three 
$40h^{-1}\dim{Mpc}$ boxes and blue squares show six independent $20h^{-1}\dim{Mpc}$ boxes. Besides a single octant outlier, the two distributions are reasonably similar (notice, that higher mean density regions actually have lower opacity, because they are reionized earlier).

Hence, we conclude that the DC mode formalism is a reasonable numerical approach to take. This test, of course, does not check against all numerical artifacts, our simulated box sizes may still be too small and some important physics may be missing from the simulation model. The test only assures us that we can meaningfully use an ensemble of independent realizations to compute statistical properties of a larger volume that we cannot, as of now, yet model numerically.
 
\bibliographystyle{apj}

\bibliography{ng-bibs/igm,ng-bibs/rei,ng-bibs/self,croc}

\begin{thebibliography}{31}
\expandafter\ifx\csname natexlab\endcsname\relax\def\natexlab#1{#1}\fi

\bibitem[{{Ba{\~n}ados} \& {Ferreira}(2014)}]{croc:banados2014}
{Ba{\~n}ados}, M. \& {Ferreira}, P.~G. 2014, Physical Review Letters, 113,
  119901

\bibitem[{{Becker} {et~al.}(2011){Becker}, {Bolton}, {Haehnelt}, \&
  {Sargent}}]{igm:bbh11}
{Becker}, G.~D., {Bolton}, J.~S., {Haehnelt}, M.~G., \& {Sargent}, W.~L.~W.
  2011, \mnras, 410, 1096

\bibitem[{{Becker} {et~al.}(2015{\natexlab{a}}){Becker}, {Bolton}, \&
  {Lidz}}]{croc:becker2015}
{Becker}, G.~D., {Bolton}, J.~S., \& {Lidz}, A. 2015{\natexlab{a}}, ArXiv
  e-prints

\bibitem[{{Becker} {et~al.}(2015{\natexlab{b}}){Becker}, {Bolton}, {Madau},
  {Pettini}, {Ryan-Weber}, \& {Venemans}}]{igm:bbm15}
{Becker}, G.~D., {Bolton}, J.~S., {Madau}, P., {Pettini}, M., {Ryan-Weber},
  E.~V., \& {Venemans}, B.~P. 2015{\natexlab{b}}, \mnras, 447, 3402

\bibitem[{{Becker} {et~al.}(2012){Becker}, {Sargent}, {Rauch}, \&
  {Carswell}}]{igm:bsr12}
{Becker}, G.~D., {Sargent}, W.~L.~W., {Rauch}, M., \& {Carswell}, R.~F. 2012,
  \apj, 744, 91

\bibitem[{{Becker} {et~al.}(2006){Becker}, {Sargent}, {Rauch}, \&
  {Simcoe}}]{igm:bsr06}
{Becker}, G.~D., {Sargent}, W.~L.~W., {Rauch}, M., \& {Simcoe}, R.~A. 2006,
  \apj, 640, 69

\bibitem[{{Becker} {et~al.}(2001){Becker}, {Fan}, {White}, {Strauss},
  {Narayanan}, {Lupton}, {Gunn}, {Annis}, {Bahcall}, {Brinkmann}, {Connolly},
  {Csabai}, {Czarapata}, {Doi}, {Heckman}, {Hennessy}, {Ivezi{\'c}}, {Knapp},
  {Lamb}, {McKay}, {Munn}, {Nash}, {Nichol}, {Pier}, {Richards}, {Schneider},
  {Stoughton}, {Szalay}, {Thakar}, \& {York}}]{igm:bfws01}
{Becker}, R.~H., {Fan}, X., {White}, R.~L., {Strauss}, M.~A., {Narayanan},
  V.~K., {Lupton}, R.~H., {Gunn}, J.~E., {Annis}, J., {Bahcall}, N.~A.,
  {Brinkmann}, J., {Connolly}, A.~J., {Csabai}, I., {Czarapata}, P.~C., {Doi},
  M., {Heckman}, T.~M., {Hennessy}, G.~S., {Ivezi{\'c}}, {\v Z}., {Knapp},
  G.~R., {Lamb}, D.~Q., {McKay}, T.~A., {Munn}, J.~A., {Nash}, T., {Nichol},
  R., {Pier}, J.~R., {Richards}, G.~T., {Schneider}, D.~P., {Stoughton}, C.,
  {Szalay}, A.~S., {Thakar}, A.~R., \& {York}, D.~G. 2001, \aj, 122, 2850

\bibitem[{{Bolton} {et~al.}(2011){Bolton}, {Becker}, {Raskutti}, {Wyithe},
  {Haehnelt}, \& {Sargent}}]{croc:bolton2011}
{Bolton}, J.~S., {Becker}, G.~D., {Raskutti}, S., {Wyithe}, J.~S.~B.,
  {Haehnelt}, M.~G., \& {Sargent}, W.~L.~W. 2011, ArXiv e-prints

\bibitem[{{Carilli} {et~al.}(2010){Carilli}, {Wang}, {Fan}, {Walter}, {Kurk},
  {Riechers}, {Wagg}, {Hennawi}, {Jiang}, {Menten}, {Bertoldi}, {Strauss}, \&
  {Cox}}]{rei:cwf10}
{Carilli}, C.~L., {Wang}, R., {Fan}, X., {Walter}, F., {Kurk}, J., {Riechers},
  D., {Wagg}, J., {Hennawi}, J., {Jiang}, L., {Menten}, K.~M., {Bertoldi}, F.,
  {Strauss}, M.~A., \& {Cox}, P. 2010, \apj, 714, 834

\bibitem[{{Chornock} {et~al.}(2013){Chornock}, {Berger}, {Fox}, {Lunnan},
  {Drout}, {Fong}, {Laskar}, \& {Roth}}]{croc:chornock2013}
{Chornock}, R., {Berger}, E., {Fox}, D.~B., {Lunnan}, R., {Drout}, M.~R.,
  {Fong}, W.-f., {Laskar}, T., \& {Roth}, K.~C. 2013, \apj, 774, 26

\bibitem[{{Fan} {et~al.}(2006){Fan}, {Strauss}, {Becker}, {White}, {Gunn},
  {Knapp}, {Richards}, {Schneider}, {Brinkmann}, \& {Fukugita}}]{igm:fsbw06}
{Fan}, X., {Strauss}, M.~A., {Becker}, R.~H., {White}, R.~L., {Gunn}, J.~E.,
  {Knapp}, G.~R., {Richards}, G.~T., {Schneider}, D.~P., {Brinkmann}, J., \&
  {Fukugita}, M. 2006, \aj, 132, 117

\bibitem[{{Gallerani} {et~al.}(2006){Gallerani}, {Choudhury}, \&
  {Ferrara}}]{igm:gcf06}
{Gallerani}, S., {Choudhury}, T.~R., \& {Ferrara}, A. 2006, \mnras, 370, 1401

\bibitem[{{Gallerani} {et~al.}(2008{\natexlab{a}}){Gallerani}, {Ferrara},
  {Fan}, \& {Choudhury}}]{croc:gallerani2008}
{Gallerani}, S., {Ferrara}, A., {Fan}, X., \& {Choudhury}, T.~R.
  2008{\natexlab{a}}, \mnras, 386, 359

\bibitem[{{Gallerani} {et~al.}(2008{\natexlab{b}}){Gallerani}, {Ferrara},
  {Fan}, \& {Choudhury}}]{igm:gff08}
---. 2008{\natexlab{b}}, \mnras, 386, 359

\bibitem[{{Gnedin}(2014)}]{ng:g14}
{Gnedin}, N.~Y. 2014, \apj, 793, 29

\bibitem[{{Gnedin}(2016)}]{ng:g16a}
---. 2016, \apj, 821, 50

\bibitem[{{Gnedin} \& {Kaurov}(2014)}]{ng:gk14}
{Gnedin}, N.~Y. \& {Kaurov}, A.~A. 2014, \apj, 793, 30

\bibitem[{{Gnedin} {et~al.}(2011){Gnedin}, {Kravtsov}, \& {Rudd}}]{ng:gkr11}
{Gnedin}, N.~Y., {Kravtsov}, A.~V., \& {Rudd}, D.~H. 2011, \apjs, 194, 46

\bibitem[{{Gunn} \& {Peterson}(1965)}]{croc:gunnpeterson}
{Gunn}, J.~E. \& {Peterson}, B.~A. 1965, \apj, 142, 1633

\bibitem[{{Jiang} {et~al.}(2015){Jiang}, {McGreer}, {Fan}, {Bian}, {Cai},
  {Cl{\'e}ment}, {Wang}, \& {Fan}}]{igm:jmf15}
{Jiang}, L., {McGreer}, I.~D., {Fan}, X., {Bian}, F., {Cai}, Z., {Cl{\'e}ment},
  B., {Wang}, R., \& {Fan}, Z. 2015, \aj, 149, 188

\bibitem[{{Jiang} {et~al.}(2016){Jiang}, {McGreer}, {Fan}, {Strauss},
  {Ba{\~n}ados}, {Becker}, {Bian}, {Farnsworth}, {Shen}, {Wang}, {Wang},
  {Wang}, {White}, {Wu}, {Wu}, {Yang}, \& {Yang}}]{igm:jmf16}
{Jiang}, L., {McGreer}, I.~D., {Fan}, X., {Strauss}, M.~A., {Ba{\~n}ados}, E.,
  {Becker}, R.~H., {Bian}, F., {Farnsworth}, K., {Shen}, Y., {Wang}, F.,
  {Wang}, R., {Wang}, S., {White}, R.~L., {Wu}, J., {Wu}, X.-B., {Yang}, J., \&
  {Yang}, Q. 2016, \apj, 833, 222

\bibitem[{{McGreer} {et~al.}(2015){McGreer}, {Mesinger}, \&
  {D'Odorico}}]{croc:mcgreer2015}
{McGreer}, I.~D., {Mesinger}, A., \& {D'Odorico}, V. 2015, \mnras, 447, 499

\bibitem[{{Mortlock} {et~al.}(2011){Mortlock}, {Warren}, {Patel}, {Venemans},
  {McMahon}, {Hewett}, {Simpson}, {Theuns}, {Rottgering}, {Kuiper}, {Bolton},
  \& {Harhnelt}}]{croc:mortlock2011}
{Mortlock}, D., {Warren}, S., {Patel}, M., {Venemans}, B., {McMahon}, R.,
  {Hewett}, P., {Simpson}, C., {Theuns}, T., {Rottgering}, H., {Kuiper}, R.,
  {Bolton}, J., \& {Harhnelt}, M. 2011, in Galaxy Formation, 88

\bibitem[{{Paschos} \& {Norman}(2005)}]{croc:paschos2005}
{Paschos}, P. \& {Norman}, M.~L. 2005, \apj, 631, 59

\bibitem[{{Pentericci} {et~al.}(2011){Pentericci}, {Fontana}, {Vanzella},
  {Castellano}, {Grazian}, {Dijkstra}, {Boutsia}, {Cristiani}, {Dickinson},
  {Giallongo}, {Giavalisco}, {Maiolino}, {Moorwood}, {Paris}, \&
  {Santini}}]{croc:pentericci2011}
{Pentericci}, L., {Fontana}, A., {Vanzella}, E., {Castellano}, M., {Grazian},
  A., {Dijkstra}, M., {Boutsia}, K., {Cristiani}, S., {Dickinson}, M.,
  {Giallongo}, E., {Giavalisco}, M., {Maiolino}, R., {Moorwood}, A., {Paris},
  D., \& {Santini}, P. 2011, \apj, 743, 132

\bibitem[{{Schenker} {et~al.}(2014){Schenker}, {Ellis}, {Konidaris}, \&
  {Stark}}]{croc:schenker2014}
{Schenker}, M.~A., {Ellis}, R.~S., {Konidaris}, N.~P., \& {Stark}, D.~P. 2014,
  \apj, 795, 20

\bibitem[{{Songaila} \& {Cowie}(2002)}]{croc:song2002}
{Songaila}, A. \& {Cowie}, L.~L. 2002, \aj, 123, 2183

\bibitem[{{Songaila} \& {Cowie}(2010)}]{igm:sc10}
---. 2010, \apj, 721, 1448

\bibitem[{{Stark} {et~al.}(2010){Stark}, {Ellis}, {Chiu}, {Ouchi}, \&
  {Bunker}}]{croc:stark2010}
{Stark}, D.~P., {Ellis}, R.~S., {Chiu}, K., {Ouchi}, M., \& {Bunker}, A. 2010,
  \mnras, 408, 1628

\bibitem[{{Tilvi} {et~al.}(2014){Tilvi}, {Papovich}, {Finkelstein}, {Long},
  {Song}, {Dickinson}, {Ferguson}, {Koekemoer}, {Giavalisco}, \&
  {Mobasher}}]{croc:tilvi2014}
{Tilvi}, V., {Papovich}, C., {Finkelstein}, S.~L., {Long}, J., {Song}, M.,
  {Dickinson}, M., {Ferguson}, H.~C., {Koekemoer}, A.~M., {Giavalisco}, M., \&
  {Mobasher}, B. 2014, \apj, 794, 5

\bibitem[{{Treu} {et~al.}(2013){Treu}, {Schmidt}, {Trenti}, {Bradley}, \&
  {Stiavelli}}]{croc:treu2013}
{Treu}, T., {Schmidt}, K.~B., {Trenti}, M., {Bradley}, L.~D., \& {Stiavelli},
  M. 2013, \apjl, 775, L29

\end{thebibliography}

\end{document}